# High-speed image reconstruction for nonlinear structured illumination microscopy

Jingxiang Zhang[1], Tianyu Zhao[1], Xiangda Fu[1], Manming Shu[1], Jiajing Yan[1], Jinxiao Chen[1], Yansheng Liang[1], Shaowei Wang[1], Ming Lei[1,2]*

[1] MOE Key Laboratory for Nonequilibrium Synthesis and Modulation of Condensed Matter, School of Physics, Xi'an Jiaotong University, Xi'an, 710049, China

[2] State Key Laboratory of Electrical Insulation and Power Equipment, Xi'an Jiaotong University. Xi'an 710049, China

*ming.lei@mail.xjtu.edu.cn

**Abstract:** By exploiting the nonlinear responses of the fluorescent probes, the spatial resolution of structured illumination microscopy(SIM) can be further increased. However, due to the complex reconstruction process, the traditional reconstruction method of nonlinear structured illumination microscopy (NL-SIM) is relatively slow, which brings a great challenge to realizing real-time display of super-resolution results. To address these issues, an accelerated NL-SIM reconstruction algorithm was developed by extending a high-speed reconstruction framework, Joint Space and Frequency Reconstruction (JSFR) to NL-SIM. We anticipate that this algorithm will facilitate NL-SIM becoming a routine tool in biomedical laboratories.

## Introduction

Super-resolution Structured Illumination Microscopy (SR-SIM) has attracted extensive interest for living cell imaging in the life sciences due to its great compatibility in high spatial resolution, fast imaging speed, low phototoxicity and photobleaching[1-10]. Using modulated sinusoidal illumination patterns and sophisticated post-processing reconstruction algorithms, linear SIM(L-SIM) achieves up to a 2-fold increase in spatial resolution compared to widefield imaging. However, the resolution of L-SIM is lower compared with STED[11] and PALM which limits its application scenarios. To further improve the spatial resolution of SIM, Gustafsson et al. developed the nonlinear structured illumination microscopy (NL-SIM)[12, 13], which was successfully applied to observe the photo stable sample with a resolution of <50nm, but NL-SIM still suffered from the grievous phototoxicity and extremely sluggish frame rates. An intriguing approach involves using the switching of chemical states to achieve saturation for addressing high illumination intensity. In this context, reversibly photo switchable fluorescent proteins (RSFPs) offer significant advantages for achieving super-resolution at much lower light intensities[14]. In particular, Li et al. exploited the spatially patterned activation with recently developed RSFPs, termed Skylan-NS to achieve 45 to 62 nanometers resolution in approximately 20 to 40 frames[15-17]. In general, NL-SIM generates high-order harmonics of the sample spatial frequency using the nonlinear responses of the fluorescent probes, further extending the spatial resolution to more than three times the traditional diffraction limits.

In the past few years, with the application of super-resolution imaging technology in living cell imaging, real-time reconstruction and display of SR-SIM images has attracted increasing interest[5,

18, 19]. These technologies allow the users to obtain almost instantaneous feedback as if using a traditional widefield microscope, which greatly improves the work efficiency. While NL-SIM collects more raw images to solve the higher frequency spectral component, it requires much longer acquisition time compared to L-SIM. Meanwhile, the mainstream reconstruction method of NL-SIM is mainly based on the algorithm proposed by Gustafsson, in which several operations such as image decomposition and reconstruction are all manipulated in frequency domain[12, 15, 20]. In addition, sophisticated analysis is required for eliminating reconstruction artifacts and balancing separated frequency components[21-24]. As a result, it typically takes several tens of seconds to obtain a single SR image, which brings a great challenge to realize real-time imaging in NL-SIM.

In our previous work, we proposed a hybrid reconstruction method termed the Joint Space and Frequency Reconstruction (JSFR) method for L-SIM[18]. This simplified protocol decomposes the SR image into the sequence of patterned illumination with weighted coefficient, and omits the complex frequency operations in the conventional Wiener-SIM. The JSFR-SIM provides an 80-fold improvement in reconstruction speed without compromising of spatial resolution.

In this paper, we develop a fast reconstruction strategy for NL-SIM by extending the JSFR concept to nonlinear SIM, termed JSFR-NL-SIM, providing a powerful tool for real-time NL-SIM. Simulation results and experimental validation show that the image quality obtained with this new method is exactly the same as that of the conventional algorithm, which however increases the reconstruction speed to 677 times. We also derive the formulae of JSFR-NL-SIM rigorously and demonstrate this increasement is able to achieve 760 times when two higher-order harmonics generated. We hope that the presented JSFR-NL-SIM method will provide a way to reconstruct NL-SIM image in real time and will facilitate the widespread application of NL-SIM in biomedical laboratories in the future.

## Methods

### Basic principle of nonlinear SIM in Fourier domain

Compared with linear SIM, NL-SIM allows us to obtain higher frequency information from the sample and achieve higher resolution by exploiting the non-linear response of the sample to excitation luminescence. Here, we take NL-SIM with patterned activation(PA NL-SIM) as an example of the following, which is based on the Wiener-SIM and reconstructs in the frequency domain. In PA NL-SIM, non-fluorescent fluorophores are first photo-activated into a fluorescent state by means of a standing wave of activation light $I_{act}$. Subsequently, by utilizing excitation light $I_{exc}$ to excite the photo-switchable fluorescent which has the same frequency and phase of the activation light. The demodulation of both standing waves allows us to obtain high-frequency information of up to 2nd order, increasing the resolution to three times to Abbe limit, as expressed by:

$$\tilde{D}_n(k) = \sum_N m_N \tilde{O}(k \pm Nw)\tilde{H}^*(k+Nw)$$
$$= [\tilde{O}(k) + \frac{2}{3}\tilde{O}(k \pm w) + \frac{1}{6}\tilde{O}(k \pm 2w)]\tilde{H}^*(k+Nw) \quad (1)$$

where $m_N$ and $\omega$ respectively denote the modulation and angular frequency, $D_i(k)$ represents the spectrum of the final SR image, $O$ indicates the separated spectrum and $H$ is the optical transfer function.

The parameter estimation algorithms for NL-SIM and L-SIM are similar, it is desirable to set the modulation amplitude manually in terms of empirical values. After determining the illumination parameters, $0, \pm 1, \pm 2$ order high frequency information can be separated and shifted to the correct position in frequency domain. Finally, super resolution image can be obtained by reshaping the spectrum using Wiener filter. Typically, the whole reconstruction workflow requires 25 raw images with 5 phases in each orientation, results in a huge amount of matrix calculation and limit the reconstruction speed.

**Principle of nonlinear SIM in spatial domain**

Previously, we have developed a high speed reconstruction workflow for L-SIM, capable of suppressing out-of-focus background while reconstructed SR image in spatial domain, termed JSFR-SIM. In this work, we explore the basic theory of the SIM(L-SIM and NL-SIM) with a more general perspective in spatial domain and extend the concept of JSFR to NL-SIM, bypassing the complex calculation of frequency domain procedures. We first deducted the general form of the NL-SIM reconstruction algorithm in spatial domain from the perspective of the image model mentioned above. All patterned illumination images $D_i(k)$ contributes to final SR image with corresponding coefficient functions $c_i(x)$, we have(details in Supplemental Materials).

$$I_{SR}(x) = \sum_{i=1}^{5} c_i(x) D_i(x) \qquad (2)$$

We assume the general phase step of NL-SIM raw data is $2\pi(i-1)/N$, and specifically set to $2\pi(i-1)/5$ in our cases.

$$\begin{bmatrix} c_1(x) \\ c_2(x) \\ c_3(x) \\ c_4(x) \\ c_5(x) \end{bmatrix} = \begin{bmatrix} 1 & 1 & 1 & \cdots & 1 \\ 1 & cos\Delta\varphi_1 & cos\Delta\varphi_2 & \cdots & cos\Delta\varphi_2 N \\ 0 & sin\Delta\varphi_1 & cos\Delta\varphi_2 & \cdots & cos\Delta\varphi_2 N \\ \cdots & \cdots & \cdots & \cdots & \cdots \\ 1 & cosN\Delta\varphi_1 & cosN\Delta\varphi_2 & \cdots & cosN\Delta\varphi_2 N \\ 0 & sinN\Delta\varphi_1 & sinN\Delta\varphi_2 & \cdots & sinN\Delta\varphi_2 N \end{bmatrix}^{-1} \begin{bmatrix} \dfrac{1}{b_0} \\ \dfrac{1}{b_1} cos(\omega x + \varphi_0) \\ \dfrac{1}{b_1} cos(\omega x + \varphi_0) \\ \dfrac{1}{b_2} cos[2(\omega x + \varphi_0)] \\ \dfrac{1}{b_2} cos[2(\omega x + \varphi_0)] \end{bmatrix} \qquad (3)$$

The coefficient functions $c_i(x)$ is only associated with illumination parameters, including wave vectors and initial phases, and can be calculated by parameter estimation algorithm.

Next, we demonstrate the connection of the reconstruction algorithm in spatial domain and the conventional Fourier domain(Supplemental Materials). In spatial domain, the final image can be represented as the linear combination of intermediate images with weighted coefficient, as equation4 (Supplemental Materials)

$$R_1(x) = \frac{e^{i\varphi_0}}{2} F^{-1}[\tilde{O}(k-w)\cdot\tilde{H}(k)] - \frac{e^{-i\varphi_0}}{2} F^{-1}[\tilde{O}(k+w)\cdot\tilde{H}(k)]$$

$$R_2(x) = \frac{e^{i\varphi_0}}{2i} F^{-1}[\tilde{O}(k-w)\cdot\tilde{H}(k)] - \frac{e^{-i\varphi_0}}{2i} F^{-1}[\tilde{O}(k+w)\cdot\tilde{H}(k)]$$

$$R_3(x) = \frac{e^{2i\varphi_0}}{2} F^{-1}[\tilde{O}(k-2w)\cdot\tilde{H}(k)] - \frac{e^{-2i\varphi_0}}{2} F^{-1}[\tilde{O}(k+2w)\cdot\tilde{H}(k)]$$

$$R_4(x) = \frac{e^{2i\varphi_0}}{2i} F^{-1}[\tilde{O}(k-2w)\cdot\tilde{H}(k)] - \frac{e^{-2i\varphi_0}}{2i} F^{-1}[\tilde{O}(k+2w)\cdot\tilde{H}(k)]$$

(4)

where $\omega$ and $\varphi_0$ are denoted as angular frequency and the initial phase of the illumination, respectively. In the Fourier domain, equation4 can be written as

$$\tilde{O}(k-w)\cdot\tilde{H}(k) = e^{-i\varphi_0} F[R_1(x)+iR_2(x)]$$
$$\tilde{O}(k+w)\cdot\tilde{H}(k) = e^{i\varphi_0} F[R_1(x)-iR_2(x)]$$
$$\tilde{O}(k-2w)\cdot\tilde{H}(k) = e^{-2i\varphi_0} F[R_3(x)+iR_4(x)]$$
$$\tilde{O}(k+2w)\cdot\tilde{H}(k) = e^{2i\varphi_0} F[R_3(x)-iR_4(x)]$$

(5)

We concluded that the connection of five separated spectrum in frequency domain $O(k)H(k)$, $O(k\pm\omega)H(k)$, $O(k\pm 2\omega)H(k)$ with the intermediate images and verify the result of the reconstruction image recovered in spatial domain is equivalent to that of the Fourier domain. Meanwhile, furthering frequency filter which is calculated in preparation procedure can be employed for eliminating out-of-focus background and the periodic honeycomb artifacts and obtaining artifact reduced SR image as ref6.

$$D_i^{'}(x) = D_i(x) \otimes F^{-1}\left\{[1-a(k)] * \tilde{H}^*(k)\right\}$$ (6)

where $a(k)$ is the attenuation function, $H^*(k)$ is the complex conjugate of the OTF. Finally, for flattening the spectrum with respect to the OTF and compensate spectrum components, an optimization functions previously employed in JSFR-AR-SIM are reused in JSFR-NL-SIM (Supplemental Materials), as shown in equation

$$I(x) = \mathfrak{I}^{-1}\left\{\mathfrak{I}\left[\sum_{d=1}^{5}\sum_{p=1}^{5} c_i(x) D_i^{'}(x)\right] \times W(k)\right\}$$ (7)

where $c_i(x)$ is the coefficient function and $W(k)$ are the optimization functions.

The workflow of JSFR-NL-SIM as shown in figure1 which can be divided into two parts: pre-calculation and main reconstruction procedures. The illumination parameters are determined with cross correlation method proposed by Gustafsson et al and able to calculate in advanced. In main reconstruction procedures, each raw image is filtered by attenuation OTF and the filtered images multiply with coefficient functions to obtain intermediate images sequentially. Finally, to reconstruct the final SR image, all intermediate images are summed and multiplied by optimization functions.

# Results

## JSFR-NL-SIM accelerates the reconstruction significantly

We demonstrated the effectiveness of the JSFR-NL-SIM algorithm by successfully increasing the resolution of wide-field image to approximately three times the Abbe limit based on the PA NL-SIM data provided by the BioSR dataset[25]. As a validation, we compared the reconstruction time of JSFR-NL-SIM with HiFi-NL-SIM as the function of image size. In this test, we abandoned the spatial-domain optimization that performed in the end of the HiFi-NL-SIM and evaluated the execution time in both CPU and GPU environment, as shown in Table 1. In general, results show that the reconstruction time of JSFR-NL-SIM has improved by a factor of 10.1- to 13.7-fold faster than HiFi-NL-SIM in the CPU environment. With an image pixel size of $512 \times 512$, the reconstruction time of the JSFR-NL-SIM is 1174.3ms, while the HiFi-NL-SIM is 15395.9ms, which is a 13.1x speed improvement in the CPU environment. Furthermore, when executed in the GPU environment, this speed improvement has reached 128-fold. In particular, when the input image size is $1024 \times 1024$, JSFR-NL-SIM is capable of 677 times the speed increase of HiFi-Nl-SIM in GPU environment.

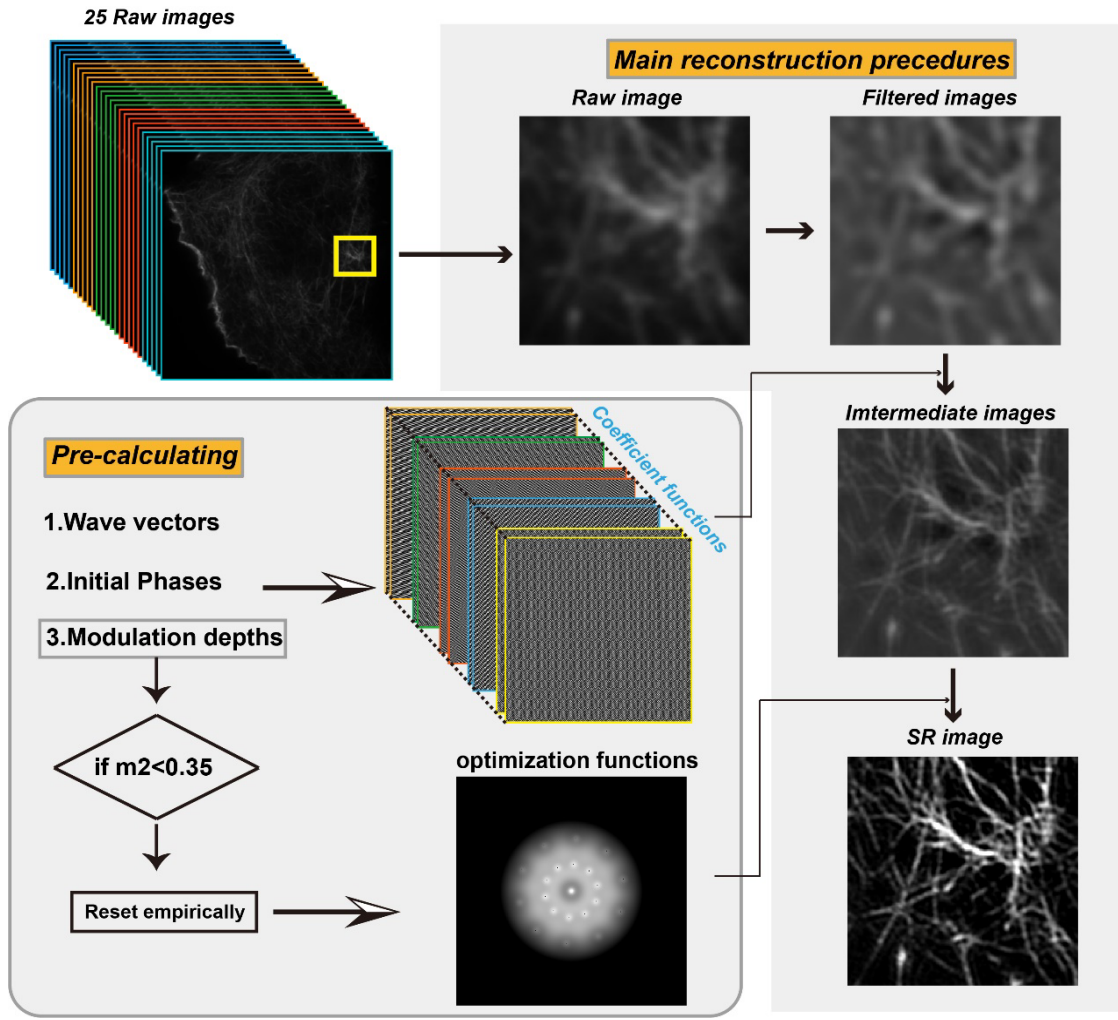

*Fig.1 Workflow of JSFR-NL-SIM can be divided into two parts: Pre-calculating and Main reconstruction procedures. In Pre-calculation, three illumination parameters including wave vectors, initial phases and modulation depths are required to calculate in advanced respectively and then the optimization function can be obtained for the final reconstruction. In main reconstruction precedures, twenty-five raw images are firstly filtered with attenuation function. Next, by summing the results of the multiplication of the coefficient functions and the filtered images, the intermediate images are obtained. Finally, the final SR image is reconstructed by filtering the optimization function. }*

## JSFR-NL-SIM produces High-Quality SR images

As a proof-of-concept, we first validate the performance of JSFR-NL-SIM capabilities to further improve the spatial resolution compared with JSFR-SIM. Resolution charts were generated by forward imaging model and the raw data were reconstructed by JSFR-SIM, JSFR-NL-SIM and HiFi-NL-SIM, respectively. As expected, the SR image obtained with JSFR-SIM improved the resolution by up to a factor of two. The resolution was further extended by adopting nonlinear SIM in which the JSFR-NL-SIM achieved the same results as the HiFi-NL-SIM.

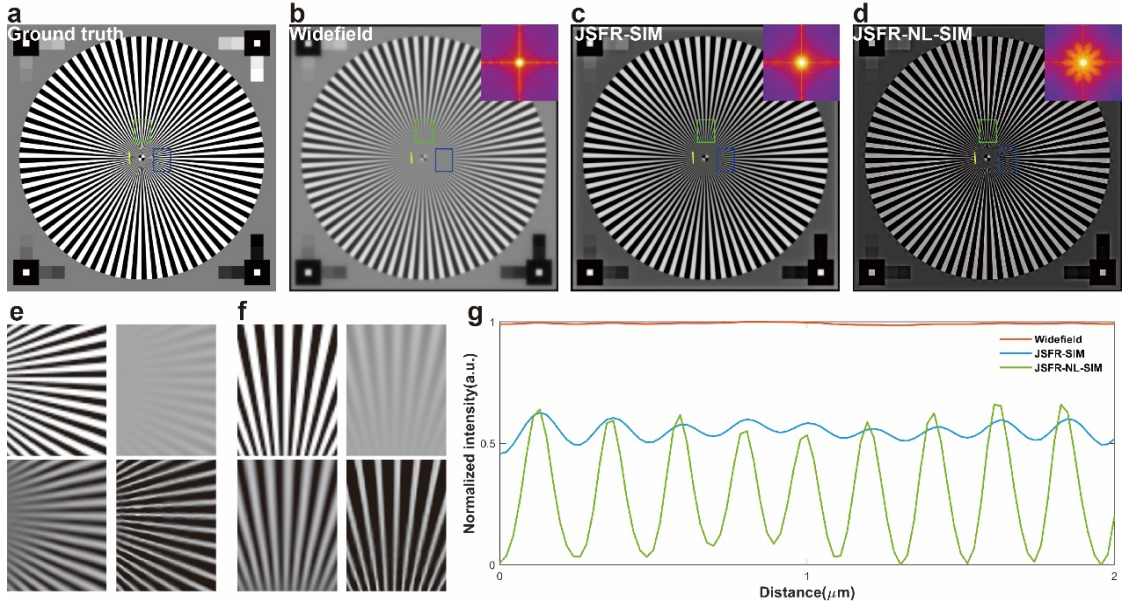

*Fig2. Simulation results to demonstrate the quality of JSFR-NL-SIM. (a) Ground truth (b) Widefield image (c) SR image reconstructed by JSFR-SIM (d) SR image reconstructed by JSFR-NL-SIM (e) and (f) are magnified*

    We also employ the PA NL-SIM data provided by BioSR[25] to further demonstrate the capability of JSFR-NL-SIM to rapidly obtain super resolved images. The SR images were reconstructed by JSFR-NL-SIM and HiFi-NL-SIM, respectively. The GT-SIM is provided by BioSR[25], as shown in the Figure1. Magnified images from white box revealed that the particulars of the F-actin can be resolved legibly by utilizing JSFR-NL-SIM and the intensity profiles was plotted along the yellow lines. Statistic results of FWHM were also obtained by randomly calculating 15 separated lines within the F-actin, in which the JSFR-NL-SIM is 75.1nm, HiFi-NL-SIM is 76.1nm, GT-SIM is 75.6nm and the wide-field image is 267.8nm, indicating that JSFR-NL-SIM is capable of resolving subtle details in reasonable quality samples.

    We do not include the parameter estimation step as the part of the reconstruction workflow with the following consideration. The illumination parameters are related with refractive-index distribution of the sample which does not change greatly in the same experiment and thus we are able to calculate those parameters in the preparation procedure[26]. Once the illumination parameters are confirmed, the optimization function can also be determined[27]. The illumination parameters including wave vector $k$, initial phases $\varphi_0$ and modulation amplitude $m$. There are several efficient and accurate parameter estimation methods currently available, including IRT-SIM[28, 29], PCA-SIM[30], normalized cross-correlation method[31-33], inverse matrix based method[34], etc, where PCA-SIM and IRT-SIM meets real-time imaging parameter estimation requirements. However, PCA-SIM and IRT-SIM are not expanding to NL-SIM at present. Therefore, we employ normalized cross-correlation method for parameter estimation and integrate it into the GPU environment for acceleration, achieving an average of 600ms to estimate complete parameters. Notably, the second harmonic modulation $m2$ obtained by means of cross-correlation in the overlap region tend to be unreliable[20, 35]. Therefore, in accordance with the theoretical value of PA NL-SIM provided by Li et al[15], if the obtained value is significantly deviant, we will set it is 1/6. Once the parameter estimation is complete, we will proceed to the reconstruction of nonlinear-SIM.

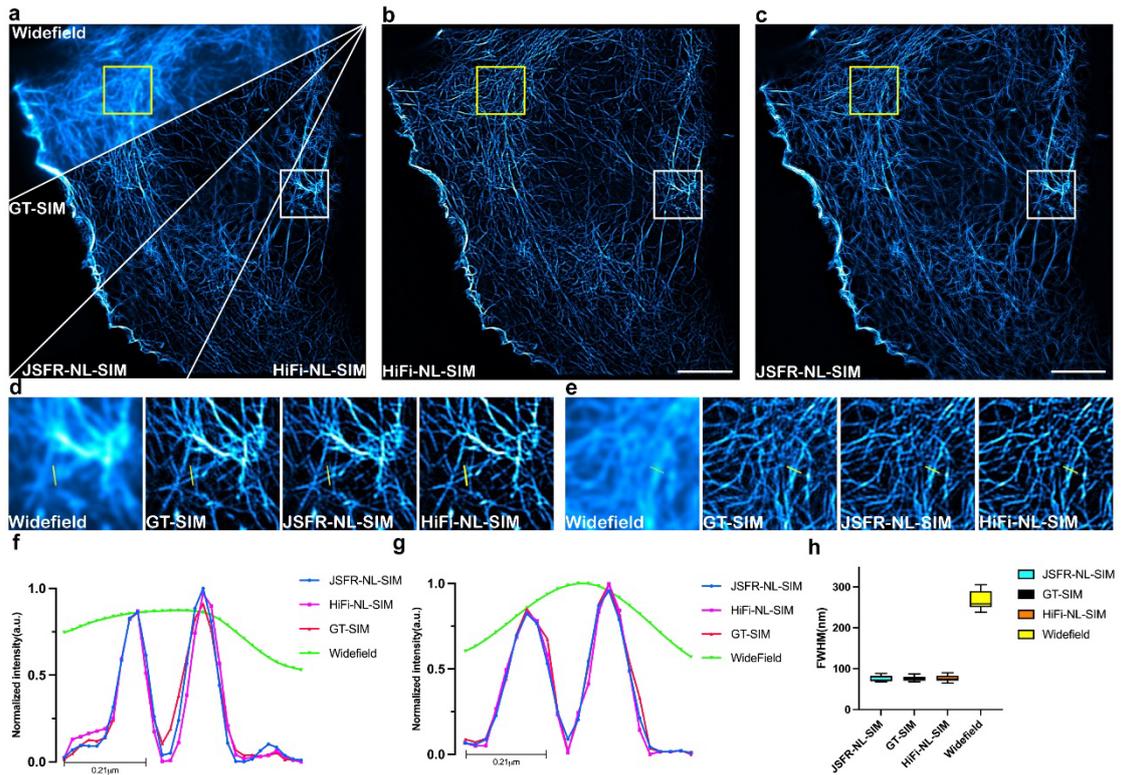

*Fig3. F-action filament from BioSR dataset[25]. (a) Pseudo-wide-field image and SR-SIM images reconstructed with JSFR-NL-SIM. (b and c) Artifact-reduced images reconstructed with HiFi- and JSFR-NL-SIM, respectively. (d and e) Close-up views of the wide-field, JSFR-SIM, HiFi-SIM, and JSFR-NL-SIM images corresponding to the yellow and blue boxes in (a), respectively. (f and g) Intensity profiles of the yellow and white lines in (d) and (e), respectively. Scale bars, 5 μm (b-c).*

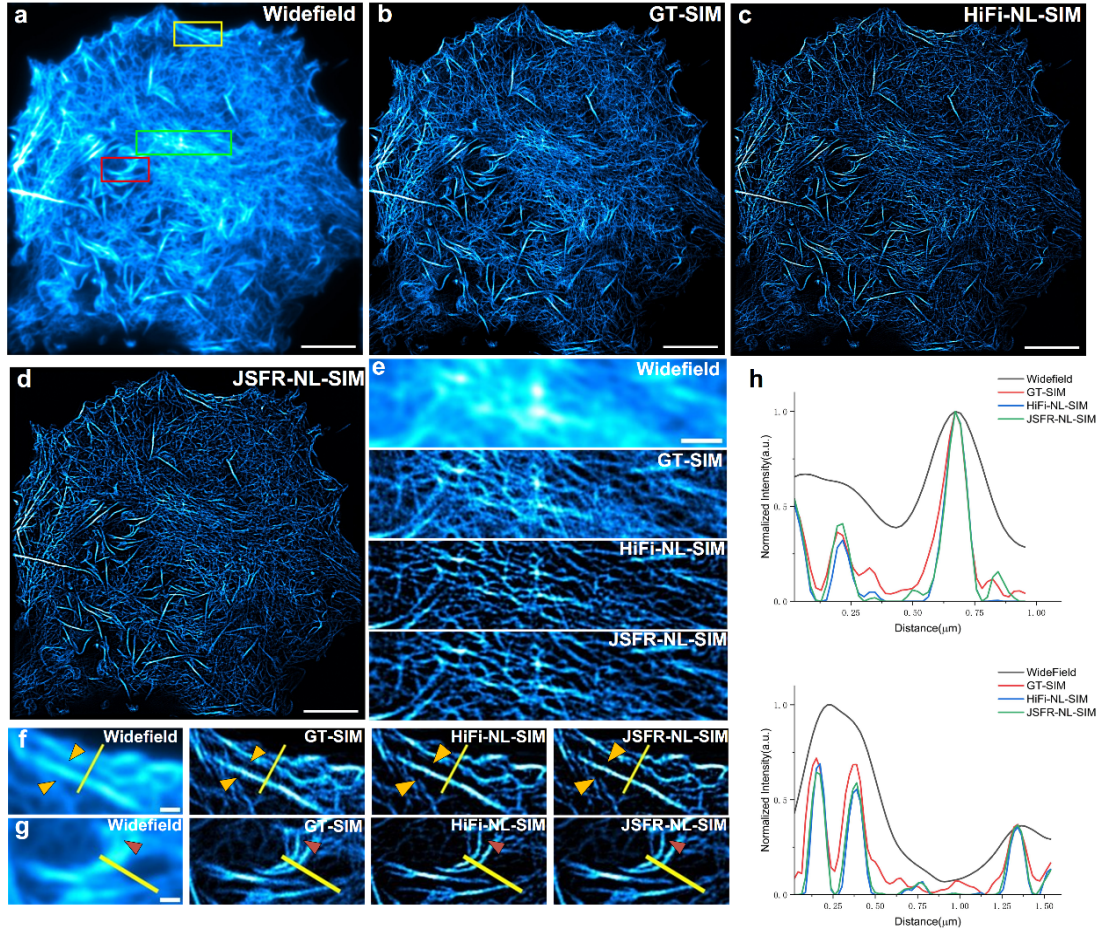

*Fig4. F-action filament from BioSR dataset[25]. (a)Widefield image. (c)-(d) SR images reconstructed by Traditional Wiener-NL-SIM, HiFi-NL-SIM and proposed JSFR-NL-SIM, respectively. (e)-(g) Zoomed-in views of Widefield, GT-SIM, HiFi-NL-SIM and JSFR-NL-SIM corresponds to green, yellow and red boxes in (a), respectively. (e) Out of focus background and filament artifacts produced by GT-SIM, but eliminated in HiFi-NL-SIM and JSFR-NL-SIM. Yellow arrows in (f) denote the artificial filament and red arrows in (g) represent the sidelobe artifact. (h) Intensity profiles of the yellow lines in (f) and (g), respectively. Scale bars, 5 μm (a-d), 1 μm (e) and 500nm(f-g).*

| Input Image Size | Output Image Size | Reconstruction time of JSFR-NL-SIM(ms) | | Reconstruction time of HiFi-NL-SIM(ms) | |
|---|---|---|---|---|---|
| | | CPU | GPU | CPU | GPU |
| 256×256 | 768×768 | 235.1±22.8 | 3.9±1.6 | 3212.5±45.7 | 165.4±14.8 |
| 512×512 | 1536×1536 | 1174.3±72.3 | 4.5±1.1 | 15395.9±92.3 | 576.6±13.3 |
| 1024×1024 | 3072×3072 | 8541.3±849.7 | 5.5±0.3 | 85885.2±11769.4 | 3725.4±83.6 |

*Table 1. Comparison of execution time reconstructed by JSFR-NL-SIM and HiFi-NL-SIM*

*a. The execution time with indicated image dimensions was evaluated using CPU processing as described in Method. Before reconstruction, the raw images are up-sampled by a factor of 3 to improve the sampling frequency. The values shown are from 100 separate processing events of each image, with the times averaged.*

*b. The comparison of the reconstruction speed was done using MATLAB (R2023a, Math Works Inc., Natick, Massachusetts, United States). The reconstruction codes for HiFi-NL-SIM and JSFR-NL-SIM (both CPU and GPU*

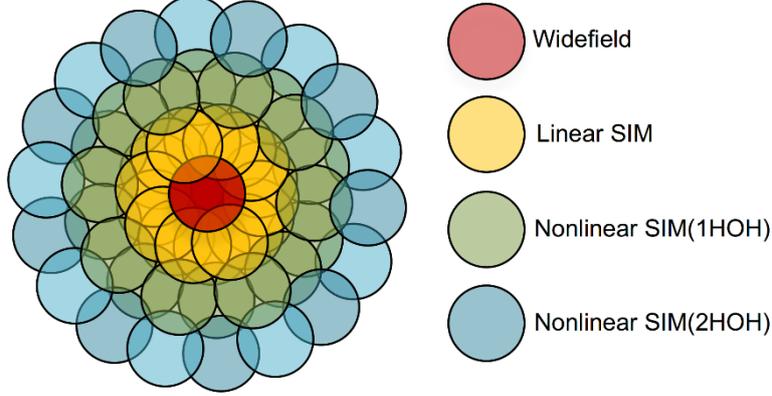

|  | Linear SIM | Nonlinear SIM(1HOH) | Nonlinear SIM(2HOH) |
|---|---|---|---|
| JSFR(GPU) | 1.5±0.2 | 4.5±1.1 | 8.7±0.7 |
| HiFi(GPU) | 81.3±1.2 | 576.6±13.3 | 6609.2±880 |

*Fig5. Comparison of execution time by JSFR and HiFi for reconstructing Linear-SIM, Nonlinear-SIM with one higher-order harmonics(1HOH) and Nonlinear-SIM with two higher-order harmonics(2HOH)(512× 512 pixel). Below the diagram is the frequency-space representation.*
*a. The execution time with indicated image dimensions was evaluated using CPU processing as described in Method. Before reconstruction, the raw images are up-sampled by a factor of 3 to improve the sampling frequency. The values shown are from 100 separate processing events of each image, with the times averaged.*
*b. The comparison of the reconstruction speed was done using MATLAB (R2023a, Math Works Inc., Natick, Massachusetts, United States). The reconstruction codes for HiFi-NL-SIM and JSFR-NL-SIM (both CPU and GPU versions) were executed on a personal computer (Intel Core i7- 13700K@2.1 GHz, DRR4 3200 MHz 64 GB, NVIDIA GeForce RTX 4070ti 12GB, Samsung 860 EVO 500GB SSD) running Windows 10.*
*c. The values in parentheses are the fold increase in processing speed of the JSFR-AR-SIM algorithm relative to HiFi-SIM in the same environment.*

## Conclusions

In this work, a rapid framework, termed JSFR-NL-SIM, was reconstructed for instantly recover SR images from wide-field images, thus robustly circumventing the long-standing and formidable challenge of real-time cell imaging. JSFR-NL-SIM was extensively tested on numerous data from open source BioSR datasets[25]. A rigorous spatial resolution criteria was also established and compared with other NL-SIM. As verified by theoretical derivation and experimental results, the final image quality and resolution obtained by JSFR-NL-SIM are consistent with the traditional NL-SIM. Meanwhile, this method enhances the reconstruction speed over 670-fold, enabling the implementation of real-time reconstruction. JSFR-NL-SIM is readily compatible with various methods, both SIM and NL-SIM, such as HiFi methods[36], Hessian denoising[37] and sparse deconvolution[38] to improve the final SR images. Integrating with the VIGOR framework proposed by Markwirth et al.[19] , this method can achieve real-time, multicolor observation with a larger FOV and shorten the latency between measurement and reconstructed image display.

Combining with the joint spectrum optimization strategy presented by G.Wen et al.[36], reconstruction artifacts in this method could be able to further eliminate. Moreover, for deep learning related methods, our method can provide faster training of datasets due to the superiority of reconstructing speed[39-41]. In particular, to detect the two higher-order harmonics, 63 raw images are required to reconstruct the nearly isotropic resolution SR image, which involves more complex calculation, further limiting the requirement for real-time imaging[20]. When dealing with such scenarios, JSFR-NL-SIM has more strengths to reconstruct SR image, as shown in Fig5. We simulate the execution time by JSFR-based method and HiFi-based method for reconstructing Linear-SIM, Nonlinear-SIM with one higher-order harmonics(1HOH) and Nonlinear-SIM with two higher-order harmonics-(2HOH) with image pixel of $512 \times 512$. Simulation results show that the execution time of the HiFi-based method increases significantly with the number of spectra to be decomposed. However, the execution time of the JSFR-based method increased linearly, all running in milliseconds. This demonstrates the superiority of JSFR-based method in terms of reconstructing time. However, due to the poor SNR, the second higher-order hamonic signal always be submerged in noise. Therefore, if more efficient fluorescence can be used in the future, JSFR-NL-SIM will provide more comprehensive assistance.

**Funding.** This work was supported in part by the National Key Research and Development Program of China (2022YFF0712500); Natural Science Foundation of China (NSFC) (Nos. 62135003, 61905189, 62205267); The Innovation Capability Support Program of Shaanxi (No. 2021TD-57); and Natural Science Basic Research Program of Shaanxi (Nos. 2020JQ-072, 2022JZ-34).